\documentclass{ifacconf}

\usepackage{natbib}        
\usepackage{graphicx}
\usepackage{amsmath}
\usepackage{amssymb}
\usepackage{tikz,pgfplots}
\pgfplotsset{compat=newest}
\usetikzlibrary{patterns}
\usepackage{algorithm}
\usepackage{algorithmic}
\usepackage{subcaption}
\usepackage{caption}
\usetikzlibrary{patterns}
\usetikzlibrary{graphs}
\usetikzlibrary{graphs.standard}
\usetikzlibrary{matrix,arrows,fit,backgrounds,mindmap,plotmarks,decorations.pathreplacing}

\newtheorem{theorem}{Theorem}
\newtheorem{definition}{Definition}
\newtheorem{lemma}{Lemma}

\newtheorem{assumption}{Assumption}

\DeclareMathOperator*{\cov}{Cov}

\DeclareMathOperator*{\adj}{adj}

\tikzstyle{sensor} = [draw, fill=blue!20, rectangle, rounded corners,
minimum height=2em, minimum width=7em]
\tikzstyle{est} = [draw, fill=orange!20, rectangle, rounded corners,
minimum height=2em, minimum width=7em]
\tikzstyle{pinstyle} = [pin edge={to-,thin,black}]
\begin{document}
\begin{frontmatter}

\title{Distributed Parameter Estimation under Gaussian Observation Noises\thanksref{footnoteinfo}} 

\thanks[footnoteinfo]{This work was supported in the part by JSPS under Grants-in-Aid for Scientific Research Grant No. 22H01508 and 21F40376.	}

\author[First]{Jiaqi Yan} 
\author[First]{Hideaki Ishii}

\address[First]{Department of Computer Science, Tokyo Institute of Technology, Japan. Emails: {jyan@sc.dis.titech.ac.jp, ishii@c.titech.ac.jp}.}

\begin{abstract}                
In this paper, we consider the problem of distributed parameter estimation in sensor networks. Each sensor makes successive observations of an unknown $d$-dimensional parameter, which might be subject to Gaussian random noises. They aim to infer true value of the unknown parameter by cooperating with each other. 
To this end, we first generalize the so-called dynamic regressor extension and mixing (DREM) algorithm to stochastic systems, with which the problem of estimating a $d$-dimensional vector parameter is transformed to that of $d$ scalar ones: one for each of the unknown parameters. 
For each of the scalar problem, an estimation scheme is given, where each sensor fuses the regressors and measurements in its in-neighborhood and updates its local estimate by using least-mean squares. Particularly, a counter is also introduced for each sensor, which prevents any (noisy) measurement from being repeatedly used such that the estimation performance will not be greatly affected by certain extreme values.
A novel excitation condition termed as \textit{local persistent excitation} (Local-PE) condition is also proposed, which relaxes the traditional persistent excitation (PE) condition and only requires that the collective signals in each sensor's in-neighborhood are sufficiently excited. With the Local-PE condition and proper step sizes, we show that the proposed estimator guarantee that each sensor infers the true parameter in mean square, even if any individual of them cannot. Numerical examples are finally provided to illustrate the established results.

\end{abstract}

\end{frontmatter}

\section{Introduction}
As a fundamental problem appearing in various applications such as signal processing, system identification and adaptive control, parameter estimation has been extensively studied in the literature  (see, e.g., \cite{goodwin2014adaptive,xie2020convergence,schizas2009distributed,chen2013distributed,cattivelli2008diffusion,yan2022distributed,yan2021resilient}). In this problem, sensors observe (partial) information of a system with an unknown (vector) parameter, and attempt to infer the true parameter through a stream of observations.

As for a single sensor, it is well known that the consistent estimation is possible only if its regressor meets certain excitation conditions \cite{goodwin2014adaptive}. Moreover, a \textit{persistent excitation} (PE) condition, which requires that the input signals are sufficiently rich such that all modes of the plant can be excited, is particularly needed to achieve an exponential convergence. 
However, in a network of multiple sensors, the PE condition may not necessarily hold at each individual sensor's side. To solve this problem, researchers leverage communication among sensors and introduce consensus algorithms into the estimators design. By doing so, weaker excitation conditions have been proposed under which the estimation task can be cooperatively fulfilled by the entire sensor network. For example, by using least-mean squares estimators, \cite{chen2013distributed} have developed a cooperative PE condition with which sensors collectively satisfy the PE condition and obtain the parameter even if no individual sensor can do so. In a recent work \cite{matveev2021diffusion}, it has also been shown that a single sensor satisfying the PE condition can lead all the others to converge to the true parameter provided that it is sufficiently connected. 

Different from the aforementioned works which focus on deterministic systems, another group of literature further considers the case where the sensors' measurements might be subject to stochastic noises (\hspace{1pt} \cite{abdolee2014diffusion,piggott2015stability,gharehshiran2013distributed,xie2018analysis,nosrati2015adaptive}). Most of them require that the regressors are generated to satisfy certain statistical independence or stationarity conditions. For example, in \cite{abdolee2014diffusion} and \cite{takahashi2010diffusion}, performance of the proposed estimators has been studied in the sensor networks where the regessors and measurement noises are independently
and identically distributed (i.i.d.) in both time and space. Despite the elegant results therein, the independence and stationarity assumptions are easily violated when the regressors are generated
from feedback systems (\cite{xie2018analysis}). In this regard, \cite{xie2018analysis} and \cite{xie2020convergence} have proposed a cooperative stochastic information condition, with which the independence
and stationarity conditions in the previous works are relaxed. 

Inspired by all these issues, in this paper, we also consider the problem of distributed parameter estimation under Gaussian observation noises. Different from the existing works, a novel excitation condition termed as \textit{local persistent excitation} (Local-PE) condition is going to be proposed, which requires the collective signals in each sensor's in-neighborhood are sufficiently excited. Note that, the Local-PE condition naturally generalizes the PE condition from a single sensor to the sensor network. However, it is strictly
weaker than the traditional PE condition, since it is possible that the regressor of each individual sensor
does not verify the PE condition, while the regressors of all sensors cooperatively
satisfy the Local-PE condition. Moreover, since no stringent conditions such as statistical independence
and stationarity are required on the regressors, the Local-PE condition is easily to meet even in the stochastic systems with feedback.

Our estimator is developed inspired by the so-called dynamic regressor extension and mixing (DREM) algorithm,
which is a new procedure to design parameter identification schemes. The DREM algorithm was first introduced in \cite{aranovskiy2017performance} and recently reviewed in \cite{ortega2020new}. However, although it reveals decent performance in simplifying the implementation, relaxing the excitation condition, and guaranteeing the asymptotic convergence, the study of this procedure only involves deterministic systems and there is no relevant work in stochastic systems. Therefore, in this paper, in order to accommodate the Gaussian noises, we first generalize the traditional DREM algorithm to stochastic settings. Then, we propose a distributed estimation scheme based on the stochastic DREM algorithm. To be specific, by leveraging DREM, we transform the problem of estimating a $d$-dimensional vector parameter to that of $d$ scalar ones: one for each of the unknown parameters. For each of the scalar problem, an estimation strategy is given, where each sensor fuses the regressors and measurements in its in-neighborhood and updates its own estimate by using least-mean squares. Moreover, notice that under the Gaussian noises, some sensors' measurements might take rather extreme values such that the estimation performance would be greatly affected. To solve this problem, in the proposed algorithm, we also introduce a counter for each sensor, which prevents any (noisy) measurement from being repeatedly used.

With the Local-PE condition and proper step sizes in making updates, the proposed estimator is proved to be efficient, in the sense that it guarantee that all sensors cooperatively fulfill the estimation task. Specifically, each of them infers the true parameter in mean square, even if any individual of them cannot.


\textit{Notations}: For a vector $v,$ we denote by $v^\prime$ its transpose. Moreover, $\mathbb{R}$ and $\mathbb{N}$ represent the sets of real numbers and natural numbers, respectively.

\section{Problem Formulation}\label{sec:form}

Let us consider a sensor network consisting of $n$ sensors. At each time instant $k$, any sensor $i\in \{1, \cdots, n\}$ receives a noisy measurement $y_{i}(k)\in\mathbb{R}$ and a $d$-dimensional regressor $\phi_i(k) \in \mathbb{R}^{d}$, which are related via the following stochastic linear regression model:
\begin{equation}\label{eqn:LRE}
y_{i}(k)=\theta^\prime \phi_i(k)+v_{i}(k), \quad k \geq 0,
\end{equation}
where $\theta\in \mathbb{R}^{m}$ is the parameter to be estimated, and $v_{i}(k)\in\mathbb{R}$ is independent and identically distributed (i.i.d.) Gaussian noise with zero mean and covariance $R_i\geq0$. Notice that $R_i$ need not to be known by any sensor.


The sensors aim to estimate $\theta$ from a stream of (noisy) measurable signals. However, in a practical network, a single sensor may not be sufficiently excited. That means the signals available at its local side are not enough to consistently estimate the parameter $\theta$. In this respect, each sensor aims to obtain an accurate estimate on $\theta$ by communicating over
a time-varying directed communication graph, which is modeled by $\mathcal{G}=(\mathcal{V},\mathcal{E}(k))$. Here, $\mathcal{V}$ is the set of sensors, and $\mathcal{E}(k)\subseteq \mathcal{V}\times\mathcal{V}$ is the set of edges. An edge from sensor $j$ to sensor $i$ is denoted by $e_{ij}(k)\in \mathcal{E}(k)$, indicating sensor $i$ can receive the information directly from sensor $j$ at time $k$. Accordingly, the sets of in-neighbors and out-neighbors of sensor $i$ are defined, respectively, as 
\begin{equation}
\begin{split}
\mathcal{N}_i^+(k)&\triangleq\{j\in \mathcal{V}|e_{ij}(k)\in \mathcal{E}(k)\},\\\mathcal{N}_i^-(k)&\triangleq\{j\in \mathcal{V}|e_{ji}(k)\in \mathcal{E}(k)\}.
\end{split}
\end{equation}
Moreover, let us denote
\begin{equation}
\mathcal{J}_i^+(k) \triangleq \mathcal{N}_i^+(k) \cup \{i\}.
\end{equation}


\section{Estimation Algorithm Design}\label{sec:algo}
This section will present the distributed estimation algorithm. Specifically, by extending the so-called \textit{dynamic regressor extension and mixing} (DREM) algorithm to stochastic settings, we decouple the $d$-dimensional estimation problem into $d$ scalar ones, each of which corresponds to an entry of the unknown vector parameter.

\subsection{Stochastic DREM}
To begin with, we shall introduce the DREM algorithm, which was first proposed in \cite{aranovskiy2017performance} and recently reviewed in \cite{ortega2020new}. Specifically, the DREM algorithm is expressed by the following variables for each sensor $i\in\mathcal{V}$:
\begin{equation}\label{eqn:definition}  
\begin{split}
\Phi_{i}(k)&\triangleq\begin{bmatrix}
\left(\phi_{i}(k)\right)^\prime \\
\left(\phi_{i}(k-1)\right)^\prime \\
\vdots \\
\left(\phi_{i}(k-d+1)\right)^\prime
\end{bmatrix}\in\mathbb{R}^{d\times d}, \\\overline{y}_{i}(k) &\triangleq\adj(\Phi_{i}(t))\left[\begin{array}{c}
y_{i}(k) \\
y_{i}(k-1) \\
\vdots \\
y_{i}(k-d+1)
\end{array}\right]\in\mathbb{R}^{d},\\
\overline\delta_{i}(k)&\triangleq \det(\Phi_{i}(k)),
\end{split}                      
\end{equation}
where we respectively denote by $\adj(\Phi_{i}(k))$ and $\det(\Phi_{i}(k))$ the adjugate matrix and determinant of matrix $\Phi_{i}(k)$.

However, note that the traditional DREM algorithm is only developed in deterministic systems (see, for example, \cite{aranovskiy2017performance,ortega2020new,pyrkin2019adaptive,yi2022conditions,matveev2021diffusion,bobtsov2022generation}). Therefore, in order to accommodate the Gaussian noises here, we should make subtle modifications on the DREM algorithm. To this end, let us define
\begin{equation}\label{eqn:noise}
\overline{v}_i(k)\triangleq \adj(\Phi_{i}(t)) \begin{bmatrix}
v_{i}(k) \\
v_{i}(k-1)\\
\vdots \\
v_{i}(k-d+1)
\end{bmatrix}.
\end{equation}
Notice that both $\overline{y}_{i}(k)$ and $\overline{v}_{i}(k)$ are vectors in the $d$-dimensional space. For simplicity, we denote by $\overline{y}^\ell_{i}(k)$ and $\overline{v}^\ell_{i}(k)$ the $\ell$-th entry of $\overline{y}_{i}(k)$ and $\overline{v}_{i}(k)$, respectively. 

We first introduce the following lemma, which extends a result in \cite{aranovskiy2017performance} to stochastic systems:

\begin{lemma}\label{lmm:Y}
	Consider the network of sensors satisfying the stochastic linear regression model \eqref{eqn:LRE}. For each $\ell\in\{1,\cdots,d\},$ it holds for any $i\in\mathcal{V}$ that
	\begin{equation}\label{eqn:scalarLRE}
	\overline{y}_i^\ell(k) = \overline\delta_{i}(k)\theta^\ell+\overline{v}_i^\ell(k),
	\end{equation}
	where $\theta^\ell$ is the $\ell$-th entry of the true parameter $\theta$.
\end{lemma}


\subsection{The proposed algorithm}
Notice that, by leveraging the stochastic DREM, we generate $d$ scalar ones as presented in \eqref{eqn:scalarLRE}: one for each of the unknown parameters. Based on it, we are ready to propose our distributed estimation algorithm.
Specifically, each sensor $i\in \mathcal{V}$ starts with the initial counter $c_i(0)=0$ and any initial estimate $\hat\theta_i(0)\in \mathbb{R}^d$. At any time $k\geq 0$, it makes an estimation as outlined in Algorithm~\ref{alg:CTA}. 

\begin{algorithm}[h!] 
	1:\: Receive $(\overline{y}_j(k), \overline\delta_j(k))$ from all in-neighboring sensors.
	
	2:\: \textbf{for} $\ell\in\{1,2,...,d\}$ \textbf{do}
	\qquad \begin{itemize}
		\item[\;] By fusing the neighboring messages, sensor $i$ updates the $\ell$-th entry of its local estimate as
		\begin{equation}\label{eqn:update}
		\begin{split}
		&\hat{\theta}_i ^\ell(k+1)=\hat{\theta}^{\ell}_{i}(k)\\&+\!\frac{\alpha(k)\Big[\sum_{j\in\mathcal{J}_i^+(k)}\delta_{j}(k)\left(\overline{y}_j^\ell(k)-\delta_{j}(k)\hat{\theta}^\ell_i(k)\right)\!\!\Big]}{\mu_i+\sum_{j\in\mathcal{J}_i^+(k)}(\delta_j(k))^2},
		\end{split}
		\end{equation}
		where $\mu_i>0$, $\alpha(k)$ is the step size to be designed later, and
		\begin{equation}\label{eqn:delta}
			\delta_{j}(k) \triangleq \begin{cases}
				\overline\delta_{j}(k), \text{ if } c_i(k) \geq d,\\
				0, \text{ otherwise,}
			\end{cases} \forall j\in\mathcal{J}_i^+(k).
		\end{equation}
	\end{itemize}
	\quad \textbf{end for}
	
	3:\: Reset the counter as
	\begin{equation}\label{eqn:counter}
		c_i(k+1) =\begin{cases}
			0, \text{ if }	\sum_{j\in\mathcal{J}_i^+(k)}(\overline\delta_{i}(k))^2 \neq 0 \text{ and } c_i(k)\geq d,\\
			c_i(k)+1, \text{ otherwise. } 
		\end{cases}
	\end{equation} 

	4:\: Transmit $(\overline{y}_i(k+1), \overline\delta_i(k+1))$ to out-neighbors. \\
	\caption{A distributed estimation algorithm under Gaussian observation noises}
	\label{alg:CTA}
\end{algorithm}

In the proposed algorithm, the sensors communicate with each other their local information on $\overline{y}_i(k)\in\mathbb{R}^d$ and $\overline\delta_i(k)\in\mathbb{R}$. Therefore, at each instant, any sensor transmits the message of size $d+1$. After that, by fusing the information within the in-neighborhood, each sensor updates its local estimate by performing \eqref{eqn:update}, which is developed based on the well-known least-mean square (LMS) algorithm (\hspace{1pt}\cite{alexander2012adaptive,lopes2008diffusion,xie2018analysis,cattivelli2008diffusion}).  
Notice that, the counter $c_i(k)$ is particularly introduced to prevent any measurement from being repeatedly used. To see this, notice that sensor $i$ updates its local estimate only when $\sum_{j\in\mathcal{J}_i^+(k)}(\delta_{j}(k))^2 \neq 0$, where measurements $\{y_j(t)\}_{j\in \mathcal{J}_i^+(k), t\in[k-d+1,k]}$ will be incorporated. After that, it resets the counter $c_i(k)$ to $0$. From \eqref{eqn:update} and \eqref{eqn:counter}, we understand that sensor $i$ can only make a new update after time $k+d$, where certain measurements after $k+1$ will be needed. Therefore, each measurement can be used at most once, which guarantees that any extreme measurement will not greatly affect the estimation performance.


\section{Performance Analysis}\label{sec:analysis}
This section is devoted to the performance analysis of Algorithm~\ref{alg:CTA}. Specifically, we would show that each sensor can infer the true parameter in mean square.

\subsection{Local persistent excitation (Local-PE) condition}
As observed from \eqref{eqn:update}, one factor that affects
convergence properties of the proposed estimator is the
determinant of the extended regressor. That is, the scalar regressor $\overline\delta_{i}(k),\;\forall i\in\mathcal{V}$. Therefore, in this subsection, we would first discuss the excitation conditions on $\overline\delta_{i}(k)$. 

In the literature, it is well known that a 
\textit{persistent excitation} (PE) condition, which guarantees the input signals to be sufficiently rich such that all modes of the plant can be
excited, is usually required to achieve an exponential convergence (\hspace{1pt}\cite{goodwin2014adaptive,aastrom2013adaptive,anderson1982exponential}). However, in the sensor network, the
PE condition may not necessarily hold at each sensor's side. To address this issue, in this paper, we will relax the PE condition
imposed on every local sensor. Instead, a new excitation condition will be proposed, which only requires
that persistently exciting signals are cooperatively generated within the neighborhood of each sensor:

\begin{definition}[Local persistent excitation (Local-PE)]\label{def:localPE}
	For any sensor $i\in\mathcal{V}$, the regressors within its in-neighborhood is said to satisfy a Local-PE condition, if there exist $\omega>0$ and a finite time $H\in\mathbb{N}_{+}$ such that 
	\begin{equation}\label{eqn:PE}
	\sum_{t=k}^{k+H-1}\Big[\sum_{j\in\mathcal{J}_i^+(k)}(\overline\delta_j(k))^2\Big] \geq \omega, \;\forall k.
	\end{equation} 
\end{definition}



\subsection{Performance analysis}
In this subsection, we shall prove that, under the Local-PE condition \eqref{eqn:PE}, the network of sensors can cooperatively fulfill
the estimation task, even if none of the individual sensors can. To see this, we first introduce assumptions that would be adopted in this paper:

\begin{assumption}\label{assup:assumptions}
	\begin{enumerate}
		\item For each sensor $i$, the regressor $\phi_i(k)$ is bounded at any time $k$. 
		\item The Local-PE condition \eqref{eqn:PE} is verified by each sensor. 
		\item The stepsize $\{\alpha(k)\}$ is monotonically non-increasing. Moreover, it satisfies that 
		\begin{equation}\label{eqn:alpha}
		0<\alpha(k)\leq 1, \;\sum_{k=0}^\infty \alpha(k) = \infty, \;\lim_{k \to \infty} \alpha(k)=0.
		\end{equation}
	\end{enumerate}
\end{assumption}

%
To prove the convergence of estimation error, let us consider any $\ell \in \{1,\cdots,d\}$. For each sensor $i\in\mathcal{V}$, we define the estimation error of it at $\ell$-th entry as
\begin{equation}
\widetilde{\theta}_i ^\ell(k) \triangleq \hat{\theta}_i ^\ell(k)-\theta^\ell.
\end{equation}
We are now ready to provide the main theorem of this paper:
\begin{theorem}\label{thm:converge}
	Consider the network of sensors satisfying the stochastic linear regression model \eqref{eqn:LRE}.
	Suppose that Assumption~\ref{assup:assumptions} holds. Then by performing Algorithm~\ref{alg:CTA}, the local estimate of each sensor converges to the true parameter in mean square. That is, for any $i\in\mathcal{V}$, it follows that
	\begin{equation}
	\lim\limits_{k\to \infty} \mathbb{E}[(\widetilde{\theta}_i^\ell(k))^2]=0.
	\end{equation}
\end{theorem}

\begin{pf}
First, notice that by \eqref{eqn:update}, for each sensor $i$, it updates the local estimate only when $c_i(k)\geq d$. Recall the definition of $\delta_{j}(k)$ in \eqref{eqn:delta}. From the Local-PE condition \eqref{eqn:PE}, there must exist $T\triangleq H+d$ such that
\begin{equation}\label{eqn:PE2}
	\sum_{t=k}^{k+T-1}\Big[\sum_{j\in\mathcal{J}_i^+(k)}(\delta_j(k))^2\Big] \geq \omega, \;\forall k.
\end{equation} 
Under this condition, it can be verified that, over any period of length $T$, sensor $i$ has at least one in-neighbor that is sufficiently excited.  That is, for any $m\in\mathbb{N}$, there exists $k\in [mT, (m+1)T-1]$ such that 
\begin{equation}
	\sum_{j\in\mathcal{J}_i^+(k)}(\delta_j(k))^2 \geq \Delta, 
\end{equation} 
where
$
\Delta \triangleq \omega/T.
$
	
Combining \eqref{eqn:scalarLRE} and \eqref{eqn:update}, the dynamics of $\widetilde{\theta}_i^\ell(k)$ is obtained as 
\begin{equation}
\begin{split}
&\widetilde{\theta}_i^\ell(k+1)=\hat{\theta}^{\ell}_{i}(k+1)-\theta^\ell\\&=\hat{\theta}^{\ell}_{i}(k)-\theta^\ell+\frac{\alpha(k)}{\mu_i+\sum_{j\in\mathcal{J}_i^+(k)}(\delta_j(k))^2}\\&\;\times\sum_{j\in\mathcal{J}_i^+(k)}\delta_{j}(k)\left((\overline{y}_j^\ell(k)-\overline\delta_{j}(k)\theta^\ell)-\delta_{j}(k)\hat{\theta}_i(k)+\overline\delta_{j}(k)\theta^\ell\right)\\&=\hat{\theta}^{\ell}_{i}(k)-\theta^\ell+\frac{\alpha(k)}{\mu_i+\sum_{j\in\mathcal{J}_i^+(k)}(\delta_j(k))^2}\\&\;\times\sum_{j\in\mathcal{J}_i^+(k)}\delta_{j}(k)\left(\overline{v}_j^\ell(k)-\delta_{j}(k)\hat{\theta}_i(k)+\overline\delta_{j}(k)\theta^\ell\right).
\end{split}
\end{equation}
Then, by \eqref{eqn:delta}, we 
can conclude
\begin{equation}\label{eqn:error}
\begin{split}
&\widetilde{\theta}_i^\ell(k+1)=\left(1-\frac{\alpha(k)\sum_{j\in\mathcal{J}_i^+(k)}(\delta_{j}(k))^2}{\mu_i+\sum_{j\in\mathcal{J}_i^+(k)}(\delta_j(k))^2}\right)\widetilde{\theta}_i^\ell(k)\\&\quad+\frac{\alpha(k)}{\mu_i+\sum_{j\in\mathcal{J}_i^+(k)}(\delta_j(k))^2}\sum_{j\in\mathcal{J}_i^+(k)}\delta_{j}(k)\overline{v}_j^\ell(k).
\end{split}
\end{equation}

We shall study both the mean and covariance of $\widetilde{\theta}_i^\ell(k)$. From \eqref{eqn:error}, it is not difficult to see that
\begin{equation}\label{eqn:mean_dyn}
\begin{split}
\mathbb{E}[\widetilde{\theta}_i^\ell(k+1)]&=\left(1-\frac{\alpha(k)\sum_{j\in\mathcal{J}_i^+(k)}(\delta_{j}(k))^2}{\mu_i+\sum_{j\in\mathcal{J}_i^+(k)}(\delta_j(k))^2}\right)\mathbb{E}[\widetilde{\theta}_i^\ell(k)] \\&=(1-\beta_i(k))\mathbb{E}[\widetilde{\theta}_i^\ell(k)],
\end{split}
\end{equation}
where
\begin{equation}\label{eqn:beta_def}
\begin{split}
\beta_i(k)\triangleq\frac{\alpha(k)\sum_{j\in\mathcal{J}_i^+(k)}(\delta_{j}(k))^2}{\mu_i+\sum_{j\in\mathcal{J}_i^+(k)}(\delta_j(k))^2}.
\end{split}
\end{equation}
It thus follows that
$
0\leq \beta_i(k)\leq \alpha(k).
$
Therefore, one knows that
$
\beta_i(k)\in[0,1].
$
Notice that when $\beta_i(k)=0$, namely, $\sum_{j\in\mathcal{J}_i^+(k)}(\delta_{j}(k))^2=0$, then $\widetilde{\theta}_i^\ell(k)$ remains unchanged. Therefore, we only focus on the case where  $\beta_i(k)\in(0,1]$.
To see this, notice that       
\begin{equation}
\begin{split}
\sum_{k=0}^{\infty} \beta_i(k) &= \sum_{m=0}^{\infty} \sum_{k=nT}^{(n+1)T-1}\beta_i(k)\\&= \sum_{m=0}^{\infty} \sum_{k=mT}^{(m+1)T-1}\frac{\alpha(k)\sum_{j\in\mathcal{J}_i^+(k)}(\delta_{j}(k))^2}{\mu_i+\sum_{j\in\mathcal{J}_i^+(k)}(\delta_j(k))^2}\\& \geq \sum_{m=0}^{\infty}\frac{\alpha((m+1)T-1)\Delta}{\mu_i+\Delta}\\&\geq \frac{\Delta}{\mu_i+\Delta}\frac{\sum_{k=T}^\infty \alpha(k)}{T}=\infty,
\end{split}
\end{equation}
where the inequalities hold due to Assumption~\ref{assup:assumptions}(3) on $\alpha(k)$. One thus concludes from \eqref{eqn:mean_dyn} that
\begin{equation}\label{eqn:mean}
\begin{split}
\lim_{k \to \infty} \mathbb{E}[\widetilde{\theta}_i^\ell(k)] &=\lim_{k \to \infty}\left(\prod_{t=0}^{k-1}(1-\beta_i(t))\right)\mathbb{E}[\widetilde{\theta}_i^\ell(0)]\\&\leq\lim_{k \to \infty} \left(\prod_{t=0}^{k-1}e^{-\beta_i(t)}\right)\mathbb{E}[\widetilde{\theta}_i^\ell(0)]\\&=e^{-\lim_{k \to \infty}\sum_{t=0}^{k-1}\beta_i(t)}\mathbb{E}[\widetilde{\theta}_i^\ell(0)]=0.
\end{split}
\end{equation}

On the other hand, we shall investigate the covariance of $\widetilde{\theta}_i^\ell(k)$. Again, we will only focus on the time instant $k$ when sensor $i$'s local estimate will be updated. As discussed previously, by introducing the counter $c_i(k)$, it follows that
$
\widetilde{\theta}_i^\ell(k) = \widetilde{\theta}_i^\ell(k-d).
$
Therefore, $\widetilde{\theta}_i^\ell(k-d)$ and $\overline{v}_j^\ell(k)$ are independent. We thus conclude from \eqref{eqn:error} that
\begin{equation}\label{eqn:cov}
\begin{split}
&\cov[\widetilde{\theta}_i^\ell(k+1)]\\=&\left(1-\frac{\alpha(k)\sum_{j\in\mathcal{J}_i^+(k)}(\delta_{j}(k))^2}{\mu_i+\sum_{j\in\mathcal{J}_i^+(k)}(\delta_j(k))^2}\right)^2\cov[\widetilde{\theta}_i^\ell(k)]\\+&\left(\frac{\alpha(k)}{\mu_i+\sum_{j\in\mathcal{J}_i^+(k)}(\delta_j(k))^2}\right)^2\sum_{j\in\mathcal{J}_i^+(k)}(\delta_{j}(k))^2\cov[\overline{v}_j^\ell(k)]\\=&\left(1-2\frac{\alpha(k)\sum_{j\in\mathcal{J}_i^+(k)}(\delta_{j}(k))^2}{\mu_i+\sum_{j\in\mathcal{J}_i^+(k)}(\delta_j(k))^2}\right. \\+&\left. \left(\frac{\alpha(k)\sum_{j\in\mathcal{J}_i^+(k)}(\delta_{j}(k))^2}{\mu_i+\sum_{j\in\mathcal{J}_i^+(k)}(\delta_j(k))^2}\right)^2\right)\cov[\widetilde{\theta}_i^\ell(k)]\\+&\left(\frac{\alpha(k)}{\mu_i+\sum_{j\in\mathcal{J}_i^+(k)}(\delta_j(k))^2}\right)^2 \sum_{j\in\mathcal{J}_i^+(k)}(\delta_j(k))^2\cov[\overline{v}_j^\ell(k)].
\end{split}
\end{equation}
Notice that
\begin{equation}
\begin{aligned}
&-\frac{\alpha(k)\sum_{j\in\mathcal{J}_i^+(k)}(\delta_{j}(k))^2}{\mu_i+\sum_{j\in\mathcal{J}_i^+(k)}(\delta_j(k))^2} + \left(\frac{\alpha(k)\sum_{j\in\mathcal{J}_i^+(k)}(\delta_{j}(k))^2}{\mu_i+\sum_{j\in\mathcal{J}_i^+(k)}(\delta_j(k))^2}\right)^2\\
&=\frac{\alpha(k)\sum_{j\in\mathcal{J}_i^+(k)}(\delta_{j}(k))^2}{\mu_i+\sum_{j\in\mathcal{J}_i^+(k)}(\delta_j(k))^2}\left( \frac{\alpha(k)\sum_{j\in\mathcal{J}_i^+(k)}(\delta_{j}(k))^2}{\mu_i+\sum_{j\in\mathcal{J}_i^+(k)}(\delta_j(k))^2}-1\right) \\&< 0,
\end{aligned}
\end{equation}
where the inequality holds as $\alpha(k)\in(0,1]$. We conclude 

\begin{equation}\label{eqn:cov2}
\begin{split}
&\cov[\widetilde{\theta}_i^\ell(k+1)]\\&<
\left(1-\frac{\alpha(k)\sum_{j\in\mathcal{J}_i^+(k)}(\delta_{j}(k))^2}{\mu_i+\sum_{j\in\mathcal{J}_i^+(k)}(\delta_j(k))^2} \right)\cov[\widetilde{\theta}_i^\ell(k)]\\&\;+\left(\frac{\alpha(k)}{\mu_i+\sum_{j\in\mathcal{J}_i^+(k)}(\delta_j(k))^2}\right)^2\sum_{j\in\mathcal{J}_i^+(k)}(\delta_j(k))^2\cov[\overline{v}_j^\ell(k)].
\end{split}
\end{equation}
For simplicity, let us define
\begin{equation}
\begin{split}
&\varepsilon_i(k) \\&\triangleq \Bigg(\frac{\alpha(k)}{\mu_i+\sum_{j\in\mathcal{J}_i^+(k)}(\delta_j(k))^2}\Bigg)^2\sum_{j\in\mathcal{J}_i^+(k)}(\delta_j(k))^2\cov[\overline{v}_j^\ell(k)]\\&\geq 0.
\end{split}
\end{equation}
We thus rewrite \eqref{eqn:cov2} as
\begin{equation}
\cov[\widetilde{\theta}_i^\ell(k+1)]< (1-\beta_i(k))\cov[\widetilde{\theta}_i^\ell(k)]+ \varepsilon_i(k).
\end{equation}
As assumed, $\phi_i(k)$ is bounded. Therefore, $\adj(\Phi_{i}(k))$ is also bounded by some constant $B<\infty$. From \eqref{eqn:noise}, for each sensor $j$, it follows that
$
\cov[\overline{v}_j^\ell(k)] \leq B^2 \max_{l\in\mathcal{V}} R_l.
$
Let $C\triangleq B^2 \max_{l\in\mathcal{V}} R_l.$ It thus holds that
\begin{equation}
\begin{split}
& \frac{\varepsilon_i(k)}{\beta_i(k)}=\frac{\alpha(k)}{\Big(\mu_i+\sum_{j\in\mathcal{J}_i^+(k)}(\delta_j(k))^2)\Big)\sum_{j\in\mathcal{J}_i^+(k)}(\delta_j(k))^2}\\&\qquad \times \sum_{j\in\mathcal{J}_i^+(k)}(\delta_j(k))^2\cov[\overline{v}_j^\ell(k)]\\
\leq & \frac{C\alpha(k)}{\Big(\mu_i+\sum_{j\in\mathcal{J}_i^+(k)}(\delta_j(k))^2)\Big)\sum_{j\in\mathcal{J}_i^+(k)}(\delta_j(k))^2}\\&\qquad \times\sum_{j\in\mathcal{J}_i^+(k)}(\delta_j(k))^2\\
=&\frac{C\alpha(k)}{\mu_i+\sum_{j\in\mathcal{J}_i^+(k)}(\delta_j(k))^2}\leq \frac{C\alpha(k)}{\mu_i}.
\end{split}
\end{equation}
As $\lim_{k \to \infty} \alpha(k)=0$, we know that
\begin{equation}
\lim_{k \to \infty} \frac{\varepsilon_i(k)}{\beta_i(k)}=0.
\end{equation}
By \cite[Lemma~A.1]{li2010consensus}, it is not difficult to conclude that
\begin{equation}
\lim_{k \to \infty} \cov[\widetilde{\theta}_i^\ell(k)] =0.
\end{equation}
Combining it with \eqref{eqn:mean}, one obtains that
\begin{equation}
\lim_{k\to \infty} \mathbb{E}[(\widetilde{\theta}_i^\ell(k))^2] = 0.
\end{equation}
Since this relationship holds for any $\ell \in\{1,\cdots,d\}$, we
finally complete the proof.\hfill$\square$
\end{pf}

In view of Theorem~\ref{thm:converge}, the convergence at each sensor side is guaranteed under the Local-PE condition. That means, through communicating with only immediate neighbors, the sensors can cooperatively fulfill the estimation task, even if any individual sensor cannot. 


\section{Simulation}\label{sec:simulation}
In this section, we will present a numerical example to demonstrate the theoretical results established in the previous sections.

\begin{figure}[!htbp]
	\centering
	\includegraphics[width=0.1\textwidth]{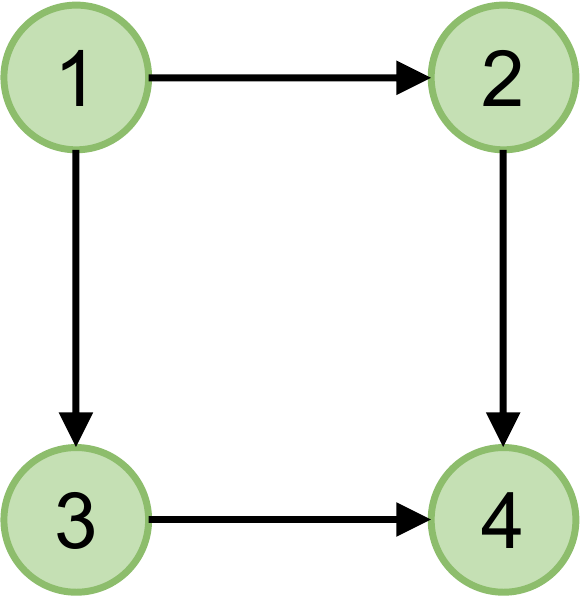}
	\caption{The communication network.}
	\label{fig:network}
\end{figure}

Let us consider the network of $4$ sensors shown in Fig~\ref{fig:network}. They aim to cooperatively estimate a $2$-dimensional parameter $\theta$ over the network. The regressor $\phi_i(k)$ for each sensor is given by
\begin{equation}
\begin{split}
\phi_1(k)&=
\begin{cases}
[1 \quad 2 ]^\prime, \text{ if } t \text{ is odd},\\
[2 \quad 3 ]^\prime, \text{ if } t \text{ is even},
\end{cases}
\\\phi_2(k)&=[
a(k) \quad 1
]^\prime,\;\phi_3(k)=[
1\quad b(k)
]^\prime,\;\phi_4(k)=[1 \quad 1 ]^\prime,
\end{split}
\end{equation}
where 
\begin{equation}
a(k)=a(k-1)+\cos\left(\frac{k\pi}{4}\right), \;b(k)=b(k-1)+\cos\left(\frac{k\pi}{2}\right),
\end{equation}
with $a(0)=1$ and $b(0)=2$. It can be checked that each sensor in the network verifies the Local-PE condition. Moreover, let the stepsize be $\alpha(k) = 0.7/k$, which meets the condition \eqref{eqn:alpha}.
We set the initial estimates of each sensors as $[0\quad 0]^\prime$ and other parameters as
$
\theta = [2.5, -1]^\prime, \;\mu_i = 0.1i,\; R_i = I, \;\forall i.
$

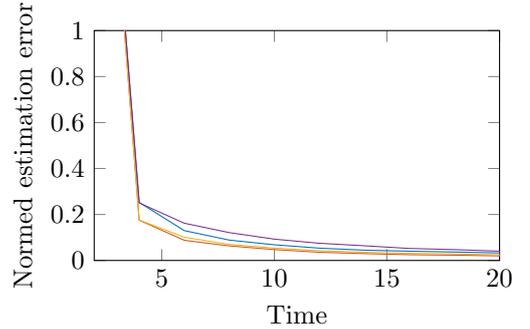
\begin{figure}[!htbp]
	\centering
%
%
\definecolor{mycolor1}{rgb}{0.00000,0.44700,0.74100}%
\definecolor{mycolor2}{rgb}{0.85000,0.32500,0.09800}%
\definecolor{mycolor3}{rgb}{0.92900,0.69400,0.12500}%
\definecolor{mycolor4}{rgb}{0.49400,0.18400,0.55600}%
\begin{tikzpicture}

\begin{axis}[%
width=2.1in,
height=1.2in,
at={(0in,0in)},
scale only axis,
xmin=1,
xmax=10,
xtick={2.5,5,7.5,10},
xticklabels={$5$,$10$,$15$,$20$},
xlabel={{Time}},
ymin=0,
ymax=1,
ylabel={{Normed estimation error}},
axis background/.style={fill=white}
]
\addplot [color=mycolor1, forget plot]
  table[row sep=crcr]{%
1	2.69258240356725\\
2	0.252465860694744\\
3	0.129825264832359\\
4	0.0886393945124605\\
5	0.0682172777095994\\
6	0.0535509789162174\\
7	0.0448115486286333\\
8	0.0403458955853084\\
9	0.0368529384674212\\
10	0.0328033714452519\\
11	0.0291778203508582\\
12	0.0259128151486051\\
13	0.0245610120681121\\
14	0.0232146599588876\\
15	0.0210739552045167\\
16	0.0195411522088204\\
17	0.0193445396369014\\
18	0.0180585887445185\\
19	0.0172657184763886\\
20	0.016973996132135\\
21	0.0163621656474095\\
22	0.0154572206364679\\
23	0.0142654271053153\\
24	0.0144633978511016\\
25	0.0144072582173494\\
26	0.0135256505275134\\
27	0.0139266181423668\\
28	0.0131559294983175\\
29	0.0125634464860409\\
30	0.0126363014363135\\
31	0.0126329017882166\\
32	0.0126676547886844\\
33	0.0125121516765135\\
34	0.0118971127596859\\
35	0.0120058933021676\\
36	0.0118605712316391\\
37	0.0120042744383859\\
38	0.0113478128360418\\
39	0.0113710194414299\\
40	0.0108824525414921\\
41	0.0109510959762961\\
42	0.0106186025482196\\
43	0.0103779846638433\\
44	0.0102444587506142\\
45	0.0100243667996751\\
46	0.00977747427342297\\
47	0.00961062739368893\\
48	0.00931199019852277\\
49	0.00907189041864474\\
50	0.00894824427976499\\
};
\addplot [color=mycolor2, forget plot]
  table[row sep=crcr]{%
1	2.69258240356725\\
2	0.17515717973663\\
3	0.0877526937171364\\
4	0.0633257982287818\\
5	0.0473295936988941\\
6	0.0356968285195119\\
7	0.0296131410531222\\
8	0.0249295157642878\\
9	0.0229818495898375\\
10	0.0197908035713157\\
11	0.0173885801126229\\
12	0.0164341910189393\\
13	0.0165903701569056\\
14	0.0162676906886727\\
15	0.0145282162955396\\
16	0.0133315564953961\\
17	0.0137188673846231\\
18	0.0124347081917033\\
19	0.0119786089585071\\
20	0.0113930564426132\\
21	0.0104942846070781\\
22	0.00909167423918167\\
23	0.00808583318104463\\
24	0.00874056435182173\\
25	0.00966974584031216\\
26	0.00880151439641734\\
27	0.00939153610958902\\
28	0.00908493146349435\\
29	0.00815909653564128\\
30	0.00837705746753641\\
31	0.00850021906618746\\
32	0.00850344567439989\\
33	0.00830771712467318\\
34	0.00809927852496654\\
35	0.00829315593323533\\
36	0.00847488445819258\\
37	0.00815420070868695\\
38	0.00754481238985503\\
39	0.00764891854834545\\
40	0.00733156084641312\\
41	0.00745549040786115\\
42	0.00721445539836726\\
43	0.00705181500559886\\
44	0.00676385473105185\\
45	0.00672191892840583\\
46	0.00667220238540426\\
47	0.00655177578392735\\
48	0.0063433438574395\\
49	0.00646850019715513\\
50	0.00629790417421743\\
};
\addplot [color=mycolor3, forget plot]
  table[row sep=crcr]{%
1	2.69258240356725\\
2	0.175835717778635\\
3	0.100990472672646\\
4	0.069941975043839\\
5	0.0538673973624648\\
6	0.0426771809765964\\
7	0.035672997935004\\
8	0.0309542728645763\\
9	0.028549714326189\\
10	0.0246716531053326\\
11	0.0211596351916766\\
12	0.0189983839496933\\
13	0.0181773852680884\\
14	0.0167319333575885\\
15	0.0151262440628538\\
16	0.0139955098825648\\
17	0.0141107419543374\\
18	0.0133232763456897\\
19	0.0130287024919931\\
20	0.0124602813256207\\
21	0.0119645728637801\\
22	0.0124962380769186\\
23	0.010610349483724\\
24	0.0119525972249121\\
25	0.0120947566745362\\
26	0.011163206964322\\
27	0.0115277176068816\\
28	0.0104799985681638\\
29	0.00997137289605056\\
30	0.0104963406580836\\
31	0.0101296112453737\\
32	0.00947390017149858\\
33	0.00936155096323267\\
34	0.00872716829713682\\
35	0.0085808225437244\\
36	0.00792524950129303\\
37	0.00818352784969091\\
38	0.007047238348653\\
39	0.00712775637580313\\
40	0.0061610368033405\\
41	0.00633860867625417\\
42	0.00584148340657369\\
43	0.00584318333814519\\
44	0.00590068238351808\\
45	0.00577121529647123\\
46	0.00570130515931057\\
47	0.00558031875877905\\
48	0.00582361879992537\\
49	0.00564380671584035\\
50	0.00553572361243235\\
};
\addplot [color=mycolor4, forget plot]
  table[row sep=crcr]{%
1	2.69258240356725\\
2	0.250348924498408\\
3	0.162041811553515\\
4	0.121138326324173\\
5	0.0927297345882038\\
6	0.0748089801210969\\
7	0.0639394016108027\\
8	0.0522082757219065\\
9	0.0472312636049996\\
10	0.0403805067354781\\
11	0.0350482272415716\\
12	0.0336490361868704\\
13	0.0334590611134806\\
14	0.0314615727416216\\
15	0.0291674201661576\\
16	0.0271760711822853\\
17	0.0269688919027579\\
18	0.0251924624065172\\
19	0.024448958428633\\
20	0.0223651119632616\\
21	0.0206487791715346\\
22	0.0204816725581136\\
23	0.0177796709339684\\
24	0.0195229953617805\\
25	0.0204911922007603\\
26	0.0188814039695647\\
27	0.018970433519743\\
28	0.0179844485689342\\
29	0.0164683476358904\\
30	0.0169967585828284\\
31	0.0159660579199418\\
32	0.0145831822280749\\
33	0.0141195604673314\\
34	0.0137410373724636\\
35	0.0133126271833777\\
36	0.0126160809825531\\
37	0.011680900314342\\
38	0.0101252149020785\\
39	0.0100060114666328\\
40	0.00882805257555881\\
41	0.00890746280467894\\
42	0.00835569090923161\\
43	0.0084937258054864\\
44	0.0082299913138635\\
45	0.0083330347405436\\
46	0.00851362849015026\\
47	0.00835525289086755\\
48	0.00872975062085192\\
49	0.00911390007867473\\
50	0.00853521343638544\\
};
\end{axis}
\end{tikzpicture}%
	\caption{Average of the Euclidean norm of estimation error of each sensor in $1000$-run Monte Carlo trials.}
	\label{fig:err}
\end{figure}

Here,
we repeat the simulation for $1000$ times with the same initial
states and parameters. The performance of Algorithm~\ref{alg:CTA} is demonstrated in Fig.~\ref{fig:err}. We can see that each sensor consistently infers the true parameter, as expected from Theorem~\ref{thm:converge}.

\section{Conclusion}\label{sec:conclude}
This paper has studied the problem of distributed parameter estimation in sensor networks, where measurements of sensors might be subject to Gaussian random noises. By generalizing the DREM algorithm to stochastic systems, a distributed estimator has been proposed, which guarantees that each sensor estimates the true parameter in mean square when the Local-PE condition and certain requirements on step sizes are met. This implies that, by working cooperatively with each other, the sensors can track a dynamic process from noisy measurements, even when none of them can fulfill the estimation task individually. Notice that in the proposed algorithm, the sensors transmit both their local (scalar) regressors and measurements.

\bibliography{reference}

\begin{thebibliography}{29}
\providecommand{\natexlab}[1]{#1}
\providecommand{\url}[1]{\texttt{#1}}
\providecommand{\urlprefix}{URL }
\expandafter\ifx\csname urlstyle\endcsname\relax
  \providecommand{\doi}[1]{doi:\discretionary{}{}{}#1}\else
  \providecommand{\doi}{doi:\discretionary{}{}{}\begingroup
  \urlstyle{rm}\Url}\fi

\bibitem[{Abdolee and Champagne(2014)}]{abdolee2014diffusion}
Abdolee, R. and Champagne, B. (2014).
\newblock Diffusion {LMS} strategies in sensor networks with noisy input data.
\newblock \emph{IEEE/ACM Transactions on Networking}, 24(1), 3--14.

\bibitem[{Alexander(2012)}]{alexander2012adaptive}
Alexander, T.S. (2012).
\newblock \emph{Adaptive {S}ignal {P}rocessing: {T}heory and {A}pplications}.
\newblock Springer.

\bibitem[{Anderson(1982)}]{anderson1982exponential}
Anderson, B.D. (1982).
\newblock Exponential convergence and persistent excitation.
\newblock In \emph{Proceedings of the 21st IEEE Conference on Decision and
  Control}, 12--17.

\bibitem[{Aranovskiy et~al.(2017)Aranovskiy, Bobtsov, Ortega, and
  Pyrkin}]{aranovskiy2017performance}
Aranovskiy, S., Bobtsov, A., Ortega, R., and Pyrkin, A. (2017).
\newblock Performance enhancement of parameter estimators via dynamic regressor
  extension and mixing.
\newblock \emph{IEEE Transactions on Automatic Control}, 62(7), 36--3550.

\bibitem[{{\AA}str{\"o}m and Wittenmark(2013)}]{aastrom2013adaptive}
{\AA}str{\"o}m, K.J. and Wittenmark, B. (2013).
\newblock \emph{Adaptive {C}ontrol}.
\newblock Courier.

\bibitem[{Bianchi et~al.(2013)Bianchi, Fort, and
  Hachem}]{bianchi2013performance}
Bianchi, P., Fort, G., and Hachem, W. (2013).
\newblock Performance of a distributed stochastic approximation algorithm.
\newblock \emph{IEEE Transactions on Information Theory}, 59(11), 7405--7418.

\bibitem[{Bobtsov et~al.(2022)Bobtsov, Yi, Ortega, and
  Astolfi}]{bobtsov2022generation}
Bobtsov, A., Yi, B., Ortega, R., and Astolfi, A. (2022).
\newblock Generation of new exciting regressors for consistent on-line
  estimation of unknown constant parameters.
\newblock \emph{IEEE Transactions on Automatic Control}.
\newblock \doi{10.1109/TAC.2022.3159568}.

\bibitem[{Cattivelli et~al.(2008)Cattivelli, Lopes, and
  Sayed}]{cattivelli2008diffusion}
Cattivelli, F.S., Lopes, C.G., and Sayed, A.H. (2008).
\newblock Diffusion recursive least-squares for distributed estimation over
  adaptive networks.
\newblock \emph{IEEE Transactions on Signal Processing}, 56(5), 1865--1877.

\bibitem[{Chen et~al.(2014)Chen, Liu, and Guo}]{chen2014stability}
Chen, C., Liu, Z., and Guo, L. (2014).
\newblock Stability of diffusion adaptive filters.
\newblock \emph{IFAC Proceedings Volumes}, 47(3), 10409--10414.

\bibitem[{Chen et~al.(2013)Chen, Wen, Hua, and Sun}]{chen2013distributed}
Chen, W., Wen, C., Hua, S., and Sun, C. (2013).
\newblock Distributed cooperative adaptive identification and control for a
  group of continuous-time systems with a cooperative {PE} condition via
  consensus.
\newblock \emph{IEEE Transactions on Automatic Control}, 59(1), 91--106.

\bibitem[{Gharehshiran et~al.(2013)Gharehshiran, Krishnamurthy, and
  Yin}]{gharehshiran2013distributed}
Gharehshiran, O.N., Krishnamurthy, V., and Yin, G. (2013).
\newblock Distributed energy-aware diffusion least mean squares: Game-theoretic
  learning.
\newblock \emph{IEEE Journal of Selected Topics in Signal Processing}, 7(5),
  821--836.

\bibitem[{Goodwin and Sin(2014)}]{goodwin2014adaptive}
Goodwin, G.C. and Sin, K.S. (2014).
\newblock \emph{Adaptive {F}iltering, {P}rediction and {C}ontrol}.
\newblock Courier.

\bibitem[{Huang and Manton(2008)}]{huang2008stochastic}
Huang, M. and Manton, J.H. (2008).
\newblock Stochastic consensus seeking with measurement noise: Convergence and
  asymptotic normality.
\newblock In \emph{Proceedings of American Control Conference}, 1337--1342.

\bibitem[{Jaakkola et~al.(1993)Jaakkola, Jordan, and
  Singh}]{jaakkola1993convergence}
Jaakkola, T., Jordan, M., and Singh, S. (1993).
\newblock Convergence of stochastic iterative dynamic programming algorithms.
\newblock \emph{Advances in Neural Information Processing Systems}, 6.

\bibitem[{Li and Zhang(2010)}]{li2010consensus}
Li, T. and Zhang, J.F. (2010).
\newblock Consensus conditions of multi-agent systems with time-varying
  topologies and stochastic communication noises.
\newblock \emph{IEEE Transactions on Automatic Control}, 55(9), 2043--2057.

\bibitem[{Lopes and Sayed(2008)}]{lopes2008diffusion}
Lopes, C.G. and Sayed, A.H. (2008).
\newblock Diffusion least-mean squares over adaptive networks: Formulation and
  performance analysis.
\newblock \emph{IEEE Transactions on Signal Processing}, 56(7), 3122--3136.

\bibitem[{Matveev et~al.(2022)Matveev, Almodarresi, Ortega, Pyrkin, and
  Xie}]{matveev2021diffusion}
Matveev, A.S., Almodarresi, M., Ortega, R., Pyrkin, A., and Xie, S. (2022).
\newblock Diffusion-based distributed parameter estimation through directed
  graphs with switching topology: Application of dynamic regressor extension
  and mixing.
\newblock \emph{IEEE Transactions on Automatic Control}, 67(8), 4256--4263.

\bibitem[{Nosrati et~al.(2015)Nosrati, Shamsi, Taheri, and
  Sedaaghi}]{nosrati2015adaptive}
Nosrati, H., Shamsi, M., Taheri, S.M., and Sedaaghi, M.H. (2015).
\newblock Adaptive networks under non-stationary conditions: formulation,
  performance analysis, and application.
\newblock \emph{IEEE Transactions on Signal Processing}, 63(16), 4300--4314.

\bibitem[{Ortega et~al.(2020)Ortega, Aranovskiy, Pyrkin, Astolfi, and
  Bobtsov}]{ortega2020new}
Ortega, R., Aranovskiy, S., Pyrkin, A.A., Astolfi, A., and Bobtsov, A.A.
  (2020).
\newblock New results on parameter estimation via dynamic regressor extension
  and mixing: Continuous and discrete-time cases.
\newblock \emph{IEEE Transactions on Automatic Control}, 66(5), 2265--2272.

\bibitem[{Piggott and Solo(2015)}]{piggott2015stability}
Piggott, M.J. and Solo, V. (2015).
\newblock Stability of distributed adaptive algorithms {II}: Diffusion
  algorithms.
\newblock In \emph{Proceedings of the 54th IEEE Conference on Decision and
  Control (CDC)}, 7428--7433. IEEE.

\bibitem[{Pu et~al.(2021)Pu, Olshevsky, and Paschalidis}]{pu2021sharp}
Pu, S., Olshevsky, A., and Paschalidis, I.C. (2021).
\newblock A sharp estimate on the transient time of distributed stochastic
  gradient descent.
\newblock \emph{IEEE Transactions on Automatic Control}.
\newblock \doi{10.1109/TAC.2021.3126253}.

\bibitem[{Pyrkin et~al.(2019)Pyrkin, Bobtsov, Ortega, Vedyakov, and
  Aranovskiy}]{pyrkin2019adaptive}
Pyrkin, A., Bobtsov, A., Ortega, R., Vedyakov, A., and Aranovskiy, S. (2019).
\newblock Adaptive state observers using dynamic regressor extension and
  mixing.
\newblock \emph{Systems \& Control Letters}, 133, 104519.

\bibitem[{Schizas et~al.(2009)Schizas, Mateos, and
  Giannakis}]{schizas2009distributed}
Schizas, I.D., Mateos, G., and Giannakis, G.B. (2009).
\newblock Distributed {LMS} for consensus-based in-network adaptive processing.
\newblock \emph{IEEE Transactions on Signal Processing}, 57(6), 2365--2382.

\bibitem[{Takahashi et~al.(2010)Takahashi, Yamada, and
  Sayed}]{takahashi2010diffusion}
Takahashi, N., Yamada, I., and Sayed, A.H. (2010).
\newblock Diffusion least-mean squares with adaptive combiners: Formulation and
  performance analysis.
\newblock \emph{IEEE Transactions on Signal Processing}, 58(9), 4795--4810.

\bibitem[{Xie and Guo(2018)}]{xie2018analysis}
Xie, S. and Guo, L. (2018).
\newblock Analysis of distributed adaptive filters based on diffusion
  strategies over sensor networks.
\newblock \emph{IEEE Transactions on Automatic Control}, 63(11), 3643--3658.

\bibitem[{Xie et~al.(2020)Xie, Zhang, and Guo}]{xie2020convergence}
Xie, S., Zhang, Y., and Guo, L. (2020).
\newblock Convergence of a distributed least squares.
\newblock \emph{IEEE Transactions on Automatic Control}, 66(10), 4952--4959.

\bibitem[{Yan et~al.(2021)Yan, Deng, and Wen}]{yan2021resilient}
Yan, J., Deng, C., and Wen, C. (2021).
\newblock Resilient output regulation in heterogeneous networked systems under
  {B}yzantine agents.
\newblock \emph{Automatica}, 133, 109872.

\bibitem[{Yan et~al.(2022)Yan, Yang, Mo, and You}]{yan2022distributed}
Yan, J., Yang, X., Mo, Y., and You, K. (2022).
\newblock A distributed implementation of steady-state {K}alman filter.
\newblock \emph{IEEE Transactions on Automatic Control}.
\newblock \doi{10.1109/TAC.2022.3175925}.

\bibitem[{Yi and Ortega(2022)}]{yi2022conditions}
Yi, B. and Ortega, R. (2022).
\newblock Conditions for convergence of dynamic regressor extension and mixing
  parameter estimators using {LTI} filters.
\newblock \emph{IEEE Transactions on Automatic Control}.
\newblock \doi{10.1109/TAC.2022.3149964}.

\end{thebibliography}

\end{document}